\documentclass{ws-mpla}

\newcommand{\ba}{\begin{array}}
\newcommand{\ea}{\end{array}}
\newcommand{\bd}{\begin{displaymath}}
\newcommand{\ed}{\end{displaymath}}
\newcommand{\be}{\begin{equation}}
\newcommand{\ee}{\end{equation}}
\newcommand{\bea}{\begin{eqnarray}}
\newcommand{\eea}{\end{eqnarray}}

\def\dis{\displaystyle}
\def\barr{\begin{array}}
\def\earr{\end{array}}
                              \def\gev{\: \rm GeV}
                              \def\tev{\: \rm TeV}
                              
                              \def\fb {\: \rm fb}
\newcommand{\beqn}{\begin{eqnarray}}
\newcommand{\eeqn}{\end{eqnarray}}

\def\ra{\rightarrow}

\def\b{\beta}

\def\q2 {q^2}

\def\r {\rightarrow}

\def\rslep {\tilde{e}_R}

\def\N10{\widetilde \chi_1^0}
\def\Cp1{\widetilde \chi_1^+}
\def\Cm1{\widetilde \chi_1^-}
\def\C1pm{\widetilde \chi_1^\pm}
\def\Ntwo{\widetilde \chi_2^0}

\def\sneu{\tilde \nu}
\def\lslep{\tilde{e}_L}

\def\mpT{p_T \hspace{-1em}/\;\:}
\def\mET{E_T \hspace{-1em}/\;\:}

\def\go{\rightarrow}
\def\goes{\longrightarrow}
\def\lsim{\:\raisebox{-0.5ex}{$\stackrel{\textstyle<}{\sim}$}\:}

\begin{document}

\markboth{Sourov Roy}
{Anomaly mediated supersymmetry breaking and its test}

%
\catchline{}{}{}{}{}
%

\title{ANOMALY MEDIATED SUPERSYMMETRY BREAKING\\
AND ITS TEST IN LINEAR COLLIDERS}

\author{\footnotesize SOUROV ROY}

\address{Helsinki Institute of Physics, P.O. Box 64 \\
FIN-00014 University of Helsinki, Finland\\
roy@pcu.helsinki.fi}

\maketitle


\begin{abstract}
Signatures of anomaly mediated supersymmetry breaking in linear
colliders are briefly reviewed after presenting an outline of the
theoretical framework. A unique and distinct feature of a large class 
of models of this type is a winolike chargino which is very closely 
degenerate in mass with the lightest neutralino. The very slow
decay of this chargino results in a heavily ionizing charged track and
one soft charged pion with a characteristic momentum distribution, 
leading to unique signals in linear colliders which are essentially 
free of background. The determination of chargino and slepton masses 
from such events is a distinctly interesting possibility.    

\end{abstract}

\ccode{PACS Nos.: 12.60.Jv,14.80.Ly}

\section{Introduction}	
          \label{sect:introd}
The Minimal Supersymmetric Standard Model (MSSM), with softly broken N=1
supersymmetry (SUSY) is perhaps the most well-motivated theory for
physics beyond the Standard Model (SM). The question then arises as to 
how SUSY is broken and why is the mechanism of SUSY breaking relevant to
experiments. The most general version of the MSSM, containing a large 
number of arbitrary SUSY breaking soft parameters, is somewhat
intractable from the viewpoint of experimental searches. However, if the 
mechanism of SUSY breaking and the way SUSY breaking is communicated to
the MSSM superfields were known, then it would be possible to  
have a much smaller set of parameters. The nature of the lightest 
supersymmetric particle (LSP) and the decay chains could then be 
established from the prediction of mass spectra and couplings from the
framework underlying SUSY breaking. Hence, the theorist's approach would
be to understand the origin of these parameters in terms of some kind of 
spontaneous SUSY breaking. 

It had been realized\cite{dimopoulos-georgi} that if spontaneous SUSY
breaking arose from MSSM fields themselves then the spectrum should obey
certain sum rules, namely $STr M_d^2 + STr M_u^2 = 0 = STr M_e^2 + STr
M_\nu^2$, generation by generation. This implies that in every family
at least some sleptons/squarks must be lighter than their partner
leptons/quarks, contrary to observation. However, this equation holds
only for renormalizable theories and at tree level. Thus, in order to
avoid this situation one can imagine that SUSY is broken dynamically in
some `hidden' sector which is singlet under the SM interactions and then
the message of SUSY breaking is communicated to the `visible' sector of
MSSM fields through some mediation mechanisms which involve either loops
or non-renormalizable interactions. The nature of the soft SUSY breaking
in the visible sector is thus determined by the mediation mechanism 
of SUSY breaking.

Several such mediation mechanisms have been discussed in 
the literature, such as supergravitational mediation at the tree 
level\cite{nilles} and gauge mediation\cite{giudice-rattazzi}, along 
with their distinctive phenomenological signatures. Anomaly mediated 
supersymmetry breaking (AMSB)\cite{randall-sundrum,giudice-luty} is 
one such possibility and has attracted a lot of attention in recent 
times. The basic idea is to convey the message of supersymmetry 
breaking to the observable sector
by the Super-Weyl anomaly. This scenario has led to the construction of
a whole class of models\cite{bagger,pomarol,kribs,katz,jack,carena,chacko,dedes,chacko1,jackjones,kaplan,weiner,campos,anisimov,luty-sundrum,harnik,wild,murakami} and many of the phenomenological implications have been discussed\cite{kribs,rat_stru_pom,feng,wells,fengmoroi,ssu,paige,tata,konar,barr,datta-huitu,shyama-ad,dkgprsr,ghosh,choi,diaz-rp,e-gamma,gamma_gamma}. 
For example,
the characteristic signatures of the minimal anomaly mediated
supersymmetry 
breaking (mAMSB) model have been studied in the context of hadronic 
colliders\cite{feng,wells,fengmoroi,ssu,paige,tata,konar,barr,datta-huitu},
as well as for high energy linear colliders, whether of the 
$e^+e^-$ type\cite{shyama-ad,dkgprsr,ghosh,choi,diaz-rp}, or $e^-\gamma$ 
colliders\cite{e-gamma} or photon-photon colliders\cite{gamma_gamma}. 
In this article we will briefly review the unique signatures of  
the mAMSB model in the context of high energy linear colliders.

The original version of the AMSB scenario involved a higher-dimensional
supergravity theory where the hidden sector and observable sector 
superfields are localized on two distinct parallel three-branes 
separated by a distance $\sim r_c$ ($r_c$ is the compactification radius) 
in the extra dimension.\cite{randall-sundrum} 
As there is no tree level coupling between the hidden sector fields and 
those in the observable sector, it had been assumed that the  
flavor changing neutral current (FCNC) processes would be suppressed
naturally. Recently, though, 
it has been shown that the physical separation between the visible and
hidden sectors in extra dimension is not sufficient to suppress the FCNC
processes except in some special cases.\cite{anisimov} However, it is 
possible to construct models of AMSB, even  
in a completely four-dimensional framework, 
that can circumvent the flavour problem.\cite{luty-sundrum,harnik} 
For the purposes of the present review, 
we would be concentrating on the minimal scenario of 
Ref. \cite{randall-sundrum}.

The AMSB scenario can be described in terms of a four-dimensional
effective theory below the compactification scale, $\mu_c (\sim
r_c^{-1})$ where only four-dimensional supergravity fields
are assumed to propagate in the bulk. It is then possible to define a 
rescaling transformation which can eliminate, from the classical 
Lagrangian, any tree-level interaction (except for the $\mu$-term in the
superpotential) connecting the supergravity fields with the visible 
sector matter fields. Such a rescaling transformation, however, is 
anomalous at the quantum level and hence supersymmetry breaking is 
communicated from the hidden sector to the visible 
sector through the loop-generated super-conformal 
anomaly.\cite{randall-sundrum} The 
supersymmetry breaking soft mass parameters for the gauginos and the
scalars 
are generated at the same order in the corresponding gauge coupling
strength. 
The analytical expressions for the scalar and gaugino masses,
in terms of the supersymmetry breaking parameters, are renormalization
group 
(RG) invariant, and, thus, can be computed at the low-energy scale 
in terms of the appropriate beta functions and anomalous dimensions. 
However, at low energies, it predicts the existence of tachyonic sleptons. 
This is the most glaring problem of the minimal AMSB scenario. 
Several solutions to this problem exist. In this review we shall consider 
mainly the minimal AMSB model wherein a constant term $m^2_0$ is added to 
all the scalar squared masses thereby making the slepton 
mass-squareds sufficiently positive.
While this may seem to be an {\em ad hoc}\ step, 
models have been constructed\cite{kaplan} that naturally lead 
to such an eventuality. A consequence is that the RG 
invariance of the expression for scalar masses is lost and hence 
one needs to consider the corresponding 
evolution down to the electroweak scale.

The minimal AMSB model has several unique features. The gravitino is 
very heavy, its mass being in the range of tens of TeV. Left and right 
selectrons and smuons are nearly mass-degenerate while the staus split
into 
two distinct mass eigenstates. But perhaps the most striking feature is 
that both the lightest supersymmetric particle (LSP) $\N10$ and 
the lighter chargino ($\C1pm$) are predominantly Winos and hence  
nearly mass-degenerate. Loop corrections as well as a small 
gaugino-Higgsino mixing at the tree level do split the 
two,\cite{wells,loop-diff} but the 
consequent mass difference is very small: $\Delta M < 1 {\rm GeV}$.
The dominant decay mode of the lighter chargino is $\C1pm \r \N10 +
\pi^\pm$ and this long-lived chargino would typically
result in  a heavily ionizing charged
track and/or a characteristic soft pion in the detector.\cite{chen-drees} 

The review is organized as follows. In Sec.2 we discuss the spectra  
in the mAMSB scenario and constraints on the parameter space 
coming from various experimental and theoretical considerations. 
In Sec.3, we first mention the results of the experimental searches of 
mAMSB scenario at LEP and then go on to discuss the signatures of mAMSB 
scenario in the context of a high energy $e^+ e^-$ linear collider. Next, 
in Sec.4 we discuss the potential of an $e^- \gamma$ collider 
and $\gamma\gamma$ collider in order to look for the signatures of AMSB. 
Finally, we conclude in Sec.5.

\section{Model Parameters and Constraints}

The minimal AMSB model is described by just three parameters (apart 
from the SM parameters, of course): the gravitino mass $m_{3/2}$, 
the common scalar mass parameter $m_0$ and $\tan\b$, the ratio 
of the two Higgs vacuum expectation values. 
In addition, there is a discrete variable, namely 
the sign of the Higgs mass term ($\mu$). As mentioned earlier, 
the soft supersymmetry breaking terms in the effective Lagrangian are then 
determined solely in terms of the gauge ($g_i$) and Yukawa ($y_a$)
couplings.

Denoting the generic beta-functions and anomalous dimension
by $\b_g(g,y) \equiv dg/dt$,
$\b_y(g,y) \equiv dy/dt$ and $\gamma(g,y) \equiv dlnZ/dt$ ($t$ being
the logarithmic scale variable) respectively, we have,
for the gaugino ($\lambda$)
masses
\be
M_\lambda = {\b_g \over g}m_{3/2},
\ee
where the appropriate gauge coupling and $\beta$-function are to be
considered. Similarly, for the trilinear soft breaking parameters, one
has
\be
A_y = {\b_y \over y}m_{3/2} \ .
\ee
The scalar masses can be symbolically written as, 
\be
\label{scalarmass}
m^2_{\tilde f} = m^2_0 - {\frac 1 4} \sum
{\left({{\partial \gamma} \over {\partial g}}
\b_g + {{\partial \gamma} \over {\partial y}} \b_y \right)} m^2_{3/2}.
\ee

\noindent The detailed expressions for the gaugino masses at the 
one-loop level and the squared masses for the Higgs and the other 
scalars at the two-loop level can be found in 
Refs. \cite{wells,ghosh,utpal}. Certain sum rules have also been derived
for the sparticle masses in the mAMSB model.\cite{sumrule}

A particularly interesting feature of the mAMSB model is that 
the ratios of the respective $U(1)$, $SU(2)$ and $SU(3)$ gaugino 
mass parameters $M_1$, $M_2$ and $M_3$, at low energies, turn out to
be
\be
|M_1| : |M_2| : |M_3| \approx 2.8 : 1: 7.1.
\label{eq:ratio}
\ee
An immediate consequence is that 
the lighter chargino $\C1pm$ and the lightest neutralino
$\N10$ are both almost exclusively a Wino and, hence, 
nearly degenerate in mass. As already discussed in the introduction 
a small mass difference is generated (the lighter chargino is always the 
heavier) though from the tree-level 
gaugino-Higgsino mixing as well as from the one-loop corrected chargino
and the neutralino mass matrices.\cite{wells,loop-diff} 
The mass splitting
has an approximate form:

\be
\begin{array}{rcl}
\Delta M \equiv m_{\tilde \chi_1^+} - m_{\tilde \chi_1^0}
         & = & \dis
        \frac{ M_W^4 \tan^2\theta_W}{(M_1 - M_2)\mu^2 } \sin^2 2\beta
\bigg [1+ {\cal O } \left(\frac{M_2}{\mu},\frac{M^2_W}{\mu M_1}
\right)\bigg ]
      \\[1.2ex]
& + & \dis \frac {\alpha M_2}{\pi\sin^2\theta_W}\bigg
[f\left(\frac{M_W^2}{M_2^2} \right)- \cos^2\theta_W f\left(\frac{M_Z^2}
{M_2^2}\right)\bigg ],
\end{array}
\label{eq:delm}
\ee
with
\be
f(x) \equiv
-\frac{x}{4}+\frac{x^2}{8}\ln(x) +\frac{1}{2}\left(1+\frac{x}{2} \right)
\sqrt{4x-x^2} \; \tan^{-1}\left(\frac{ (1-x) \sqrt{4x-x^2}}
                                     { 3 x - x^2} \right) \ .
\ee
For the range of $m_0$ and $m_{3/2}$ that we would be 
considering throughout this review,  $\Delta M \lsim$ 500 MeV. 
And, for very large $M_2$, the mass difference
reaches an asymptotic value of $\approx$ 165 MeV. 

To determine the parameter space allowed to the theory, 
several experimental constraints need to be considered, the most 
important ones being:
\begin{itemize}
\item $\tilde \chi_1^0$ must be the LSP;
\item $m_{{\tilde \tau}_1} > 82 $ GeV,\cite{delphi}
\item  $m_{\C1pm} >$ 86 GeV, when this chargino almost degenerate with 
the lightest neutralino.\cite{aleph}
\end{itemize}
The last constraint serves to rule out relatively low values of $m_{3/2}$ 
{\em irrespective} of the value of $m_0$. 
A detailed discussion of these issues can be found in Ref.\cite{utpal}. 

One must also consider the 
constraints\cite{fengmoroi,utpal,g-22.61,g-22.63} on the mAMSB model
parameters from the recent measurement of the muon anomalous magnetic
moment $(g_\mu -2)$. It must be borne in mind, though, that the
calculation of the SM contribution to $(g_\mu -2)$ is beset with many 
theoretical uncertainties. Constraints are also 
derivable\cite{fengmoroi,g-22.61,g-22.63} from the 
measurement of the rare decay rate $\Gamma(B \r X_s \gamma)$. 
Additional bounds may exist if one 
demands that the electroweak vacuum corresponds to the
global minimum of the scalar potential.\cite{samanta,gabrielli} 

\section{AMSB Signals at $e^+e^-$ Collider}

The search for nearly degenerate lighter chargino and the lightest
neutralino (lightest neutralino being the LSP) has been performed by all
the four experiments at LEP2 and bounds have been derived on the lighter
chargino mass at 95$\%$ C.L. from the signal which consists of a single high
$p_T$ ISR photon plus a large amount of missing 
energy.\cite{aleph,aran,alderweireld} 

\subsection{Signals of AMSB at high energy $e^+e^-$ linear collider}
\label{eposcol}

The viability of the channel $e^+e^- \goes \gamma \Cp1 \Cm1$ has also
been studied at the Next Linear Collider (NLC) with $\sqrt s$ = 500 GeV
within the framework of mAMSB models where the energetic photon and
missing energy from the neutralinos in the final state have been used to
trigger the event.\cite{shyama-ad} It has been shown that with 
suitably chosen kinematical cuts to reduce the backgrounds one could 
observe hundreds of background free events with $\cal L$ = 50 $fb^{-1}$
in a large region of the allowed parameter space. The track length of 
the decaying chargino and the impact parameter of the soft pion emitted 
from the decay of the lighter chargino have also been analyzed in order 
to see whether these characteristics can be used to make the signal 
background free. The chargino mass can be determined from kinematics and
the sneutrino mass can be determined approximately from the cross
section and these information could possibly be used to distinguish the 
mAMSB models from other models of nearly mass-degenerate lighter chargino 
and lightest neutralino.  

One could also look at pair production processes such as $e^+ e^-
\rightarrow \lslep^\pm \lslep^\mp$, $\rslep^\pm \rslep^\mp$, 
$\lslep^\pm \rslep^\mp$, $\sneu \bar {\sneu}$, $\N10 \Ntwo$, $\Ntwo\Ntwo$
at a CM energy $\sqrt s$ = 1 TeV.\cite{dkgprsr,ghosh} The signals
analyzed comprise multiple fast charged leptons any of which can be used
as a trigger plus heavily ionizing charged tracks from the chargino
and/or soft charged pions with characteristic momentum distribution plus
a large amount of missing energy. 

Depending on various parametric choices, the mAMSB sparticle spectrum 
has been broadly classified into two categories with the following 
mass orderings (sparticle symbols standing for the sparticle masses):

\begin{itemize}
\item Spectrum A: $\N10(\approx \C1pm)
< \tilde \nu < {\tilde e}_R (\approx {\tilde e}_L) < \Ntwo$
\item Spectrum B: $\N10(\approx \C1pm)
< \Ntwo < \tilde \nu < {\tilde e}_R (\approx {\tilde e}_L).$
\end{itemize}

Also, for large values of $\tan\beta$, there exists
a region where the mass of $\Ntwo$ is between that of the lighter stau
$\tilde\tau_1$ and the first generation sleptons. This variant, which
we call Spectrum B1, is characterized by
\begin{itemize}
\item Spectrum B1: $\N10(\approx \C1pm)
< \tilde\tau_1< \Ntwo < \tilde \nu < {\tilde e}_R (\approx {\tilde
e}_L).$
\end{itemize}
For most of the decay modes, the spectra B and B1 have identical
behaviour, the only exception is the
decay of $\Ntwo$, which is discussed in section \ref{decay-spec}. Staus
and smuons are not considered for production here. 

In the next section we shall enumerate all possible decay channels of
the low-lying sparticles $\C1pm$, $\Ntwo$, $\lslep$, $\rslep$ and 
$\sneu$ that result in one or more leptons plus at least one soft 
charged pion accompanied by missing energy. The selection criteria 
and conventions can be found in Ref. \cite{ghosh}.
   
\subsubsection{Decay cascade and production processes for spectrum A and
spectrum B}
\label{decay-spec}
For Spectrum A, the allowed decay channels are as follows:
\begin{enumerate}
\item $\Ntwo \go \nu \sneu,~~~\ell_L^\pm {\tilde
\ell}_L^\mp,~~~\ell_R^\pm
{\tilde \ell}_R^\mp$.
\item $\lslep \go e \N10,~~\nu_e \C1pm$.
\item $\rslep \go e \widetilde \chi^{0*}_2 \go e\nu\sneu$. Note that
$\rslep$
must have a three-body decay since $\N10$ has a vanishing bino
component.
Also, the virtual $\widetilde \chi^{0*}_2$ goes into the $\nu\sneu$
channel
rather than the $\ell_L^\pm {\tilde \ell}_L^\mp$ channel since
${\tilde \ell}_L$ is never lighter than $\rslep$ in minimal AMSB.
However, the $\tau\tilde \tau_1$ channel is not considered 
explicitly since we do not look at final states with $\tau$'s.
\item $\sneu \go \nu \N10,~~\ell^\mp \C1pm$.
\end{enumerate}

This immediately predicts the decay cascade for low-lying sparticles.
As already pointed out, $\C1pm$ results in a heavily ionizing charged
track $X_D$ and/or a soft charged pion\footnote{Henceforth $\pi$ will be
used to denote $X_D$ and/or charged soft pion.}.
Thus, the end
products of various sleptons and $\Ntwo$ are
\begin{enumerate}
\item $\sneu \go \ell^\pm \pi^\mp \mET$ (it can have a completely
invisible
mode $\sneu\go\nu\N10$ and thus can act as a virtual LSP).
\item $\lslep \go e\mET,~\pi\mET$.
\item $\rslep \go e\mET,~e\ell^\pm\pi^\mp \mET$.
\item $\Ntwo \go \ell^\pm \pi^\mp \mET,~\ell^+\ell^- \mET,~
\ell_1^+\ell_1^-\ell_2^\pm\pi^\mp \mET$ ($\ell_1,\ell_2 = e,\mu$).
\end{enumerate}

\noindent The allowed decay channels for Spectrum B are as follows:
\begin{enumerate}
\item $\lslep \go e \N10,~~e \Ntwo,~~\nu_e \C1pm$.
\item $\rslep \go e \Ntwo$. Thus $\rslep$ has a more prompt decay in
spectrum B than in spectrum A.
\item $\sneu \go \nu \N10,~~\nu\Ntwo,~~\ell^\mp \C1pm$.
\item The dominant decay modes of $\Ntwo$ are: $\Ntwo\go \N10 h$, $\Ntwo\go
\N10 Z$, $\Ntwo\go \C1pm W^{\mp}$, where $h$ is the lightest CP-even
Higgs scalar. 

For Spectrum B1 (see Section \ref{eposcol}), the 
two body channel
$\Ntwo\go\tau\tilde \tau_1$ opens up, and the branching ratios of all
the
above-mentioned two-body channels get suppressed. 
\end{enumerate}

The decay cascades for sleptons and gauginos are:
\begin{enumerate}
\item $\Ntwo \go \ell^\pm \pi^\mp \mET,~\ell^+\ell^- \mET$. $\Ntwo$ also
has a virtual LSP mode $\Ntwo\go\nu\nu\N10$. The leptons come from the
decay of $W$ and $Z$; $h$ decays dominantly into $b\bar b$ and
$\tau\bar\tau$ which we do not discuss here. 
\item $\sneu \go \ell^\pm \pi^\mp \mET,~\ell^+\ell^- \mET$ (and the
virtual
LSP mode listed for spectrum A).
\item $\lslep \go e\mET,~\pi\mET,~e\ell^+\ell^- \mET,
~e\ell^\pm\pi^\mp \mET$.
\item $\rslep \go e\mET,~e\ell^+\ell^- \mET,~e\ell^\pm\pi^\mp \mET$.
\end{enumerate}

\begin{table}[h]
\tbl{Possible one or multilepton signal with one soft pion.
All possible combinations of leptonic flavours are to be taken into
account where the flavour is not shown explicitly. This table is taken
from Ref.\protect\cite{ghosh}.}
{\begin{tabular}{@{}ccc@{}} \toprule
{\bf Spectrum} & {\bf Signals} & {\bf Parent Channels} \\ \colrule
 & $e~\pi$ & $\sneu \bar {\sneu},~~\lslep \lslep,~~\lslep
\rslep,~~\N10 \Ntwo,~~\Ntwo\Ntwo$ \\
 & $\mu~\pi$ & $\sneu \bar {\sneu},~~\N10 \Ntwo,~~ \Ntwo\Ntwo$ \\
{\bf A} & $e~e~\ell~\pi$ & $\rslep \rslep,~~\lslep \rslep,~~\N10
\Ntwo,~~
\Ntwo\Ntwo$ \\
 & $\mu~\mu~\ell~\pi$ & $\N10 \Ntwo,~~
\Ntwo\Ntwo$ \\
 & $\ell_1~\ell_1~\ell_2~\ell_2~\ell_3~\pi$ & $\Ntwo\Ntwo$
($\ell_{1,2,3}=e,~\mu$) \\ \colrule
 & $e~\pi$ & $\sneu \bar {\sneu},~~\lslep \lslep,~~\lslep\rslep,~~
\N10 \Ntwo,~~\Ntwo\Ntwo$ \\
 & $\mu~\pi$ & $\sneu \bar {\sneu},~~\lslep\lslep,~~\N10 \Ntwo,~~
\Ntwo\Ntwo$ \\{\bf B} & $e~\ell_1~\ell_2~\pi$ & $\rslep \rslep,~~\lslep
\rslep,~~
  \lslep \lslep,~~ \sneu \bar {\sneu},~~\Ntwo\Ntwo$ \\
 &                       & ($\ell_{1,2}=e,~\mu$)\\
 & $\mu~\mu~\mu~\pi$ & $\Ntwo\Ntwo,~~ \sneu \bar {\sneu}$ \\
 & $e~e~\ell_1~\ell_1~\ell_2~\pi$ & $\lslep \lslep,~~\rslep \rslep,
~\lslep \rslep$ ($\ell_{1,2}=e,~\mu$)\\ \botrule
\end{tabular}}
\end{table}

A list of all possible final states and their parent sparticles, as
discussed above, for both Spectra A and B, is given in Table 1 for one
pion
channels. A similar list for two pion channels can be seen in Ref.
\cite{ghosh}. In the next section, we discuss some of the one-pion 
signals and the dilepton plus dipion signal, in detail, but let us note 
two key features right at this point.

\begin{itemize}
\item One can sometimes have the same signal for Spectrum A or B;
however,
their sources are different. This means that the production
cross-section
and different distributions will also vary from one spectrum to the
other;
this may help discriminate between them. A useful option may be to use
one
polarized beam when some of the channels would be altogether ruled out.
\item Three charged lepton plus one soft pion ($3\ell 1\pi$) signals are
interesting in their own right. Consider the $3\mu 1\pi$ signal. For
Spectrum B, two opposite sign muons must have their invariant mass
peaked at
$m_Z$, while no such compulsion exists for Spectrum A. This can serve as
a useful discriminator between these two options. A discussion regarding
additional features can be seen in Ref. \cite{ghosh}.
\end{itemize}

\subsubsection{Some numerical results}

Cross sections for the production of various two-sparticle combinations
have been calculated at an $e^+e^-$ CM energy of 1 TeV for two values
of $\tan\beta$, namely, 10 and 30, for $\mu > 0$. These cross sections
were multiplied by the appropriate branching fractions of the
corresponding decay channels to get the final states described below.
The selection cuts used can be found in Ref. \cite{ghosh}. 

\begin{table}[h]
\tbl{Selected parameter points with $\mu> 0 $ for computed cross
sections.}
{\begin{tabular}{@{}ccccc@{}} \toprule
Spectrum & Parameter Set & $m_0$ (GeV) & $m_{3/2}$ (TeV) & $\tan\beta$
\\ \colrule
 & (a) & 340 & 44 &  10 \\
 & (b) & 350 & 42 &  10 \\
 & (c) & 360 & 39 &  10 \\
{\bf A} & (d) & 380 & 46 &  30 \\
 & (e) & 410 & 44 &  30 \\
 & (f) & 450 & 47 &  30 \\ \botrule
\end{tabular}}
\end{table}

Here, we shall display numerical
results for only a selected subset of the final states listed in section
3.1.1 and the dilepton plus dipion signal - mainly to get an idea of 
signal strengths. Specifically, let us
choose the final states $e~\pi~\mET$, $e~e~\mu~\pi~\mET$,
$e~e~\pi~\pi~\mET$. 
In Table 2, we have selected some AMSB parameter points and 
in Table 3, numbers for the cross sections in the three channels
mentioned above are displayed for spectrum A 
with the parameter points as selected in Table 2. The individual
processes have widely different contributions to these channels because
of the fact that their individual production cross sections and branching
ratios in the cascade decays are highly parameter-dependent. These 
signals are essentially background free and the detailed discussion
regarding the possible backgrounds can be found in Ref. \cite{ghosh}. We 
can see from table (3) that even in the worst cases of the signal 
cross-sections one could observe 15850, 3 and 4330 signal events in the
$e\pi+\mET$, $ ee\mu \pi+\mET $ and $ ee\pi \pi + \mET$ channels
respectively from Spectrum A assuming an integrated luminosity 
of $500~{\rm fb^{-1}}$. Similar numbers for spectrum B can be found in
Ref. \cite{ghosh}.

\begin{table}[h]
\tbl{Some selected signals in Spectrum A for sample choices of
parameters in Table 3. The contributions from different sources are also 
shown in the Table. Here, PS stands for Parameter Set.}
{\begin{tabular}{@{}ccccccccc@{}} \toprule
Signal & PS & \multicolumn{7}{c}{Cross Sections (fb)} \\
 & & $ \sneu \bar {\sneu} $ & $ \lslep {\bar {\tilde e}}_L $  & $\rslep 
{\bar {\tilde e}}_R$  & $\lslep {\bar {\tilde e}}_R + \rslep {\bar
{\tilde e}}_L$ & $\N10 {\Ntwo}$ & $\Ntwo {\Ntwo}$  & $ \rm{Total}$\\
\colrule
    & $a$               & 40.27 & 46.7 & - &0.00029 &2.46  & 0.118  & 
89.54\\ 
    & $b$               &40.94 & 45.09 &- & 0.000121 &2.48 &0.14 &88.65  \\ 
$ e \pi + \mET $ & $c$  &43.03 & 44.44 &- &$2.55 \times 10^{-5}$ &2.14 &0.13  & 89.74 \\ 
    & $d$               &30.17 & 31.63 &- &$3.24\times 10^{-8} $ &1.74 &0.032 &  63.57  \\ 
    & $e$               &26.4 & 24.33 &- & 0.0&1.35 &0.011 & 52.09 \\ 
    & $f$               &17.28& 13.43 &- & 0.0&0.99& 0.003 & 31.70 \\ 
\colrule
               & $a$ &- &- & $1.36 \times 10^{-4}$&0.010&1.44 &0.159 & 
1.61\\ 
               & $b$ &- &- &$3.65 \times 10^{-4}$ &0.012 &1.32&0.174 & 
1.50\\
$ e e \mu \pi + \mET $ & $c$ &- &- &0.00 & 0.018&1.19 &0.116 &1.32  \\ 
               & $d$ &- &- &0.00 &$2.3\times 10^{-5}$&0.014 & 0.033 &0.047  
\\ 
               & $e$ &- &- &0.00 &$4.15 \times 10^{-5}$&0.011 & 0.008 &
0.019\\ 
               & $f$ &- &- &0.00 &$2.02 \times 10^{-5}$& 0.006& 0.001 &
0.007 \\ 
\colrule
               & $a$ &24.21 &- &-&0.014 &-& 0.0511&24.27  \\ 
               & $b$ &24.94 &- &-&0.016 &-&0.0648 &25.02    \\ 
$ e e \pi \pi + \mET $ & $c$& 27.66 &- &-&0.026&-&0.0604 &27.74 \\ 
               & $d$ &16.45 &- &-&$2.7\times 10^{-5}$ &-&0.0119&16.46  \\ 
               & $e$ &14.62 &- &-&$5.04\times 10^{-5}$&-&0.0044&14.62   \\ 
               & $f$ &8.66 &- &-&$2.41\times 10^{-5}$ &-&0.000972&8.66  \\ 
\botrule
\end{tabular}}
\end{table}

\begin{figure}[th]
\vspace*{4pt}
\centerline{\psfig{file=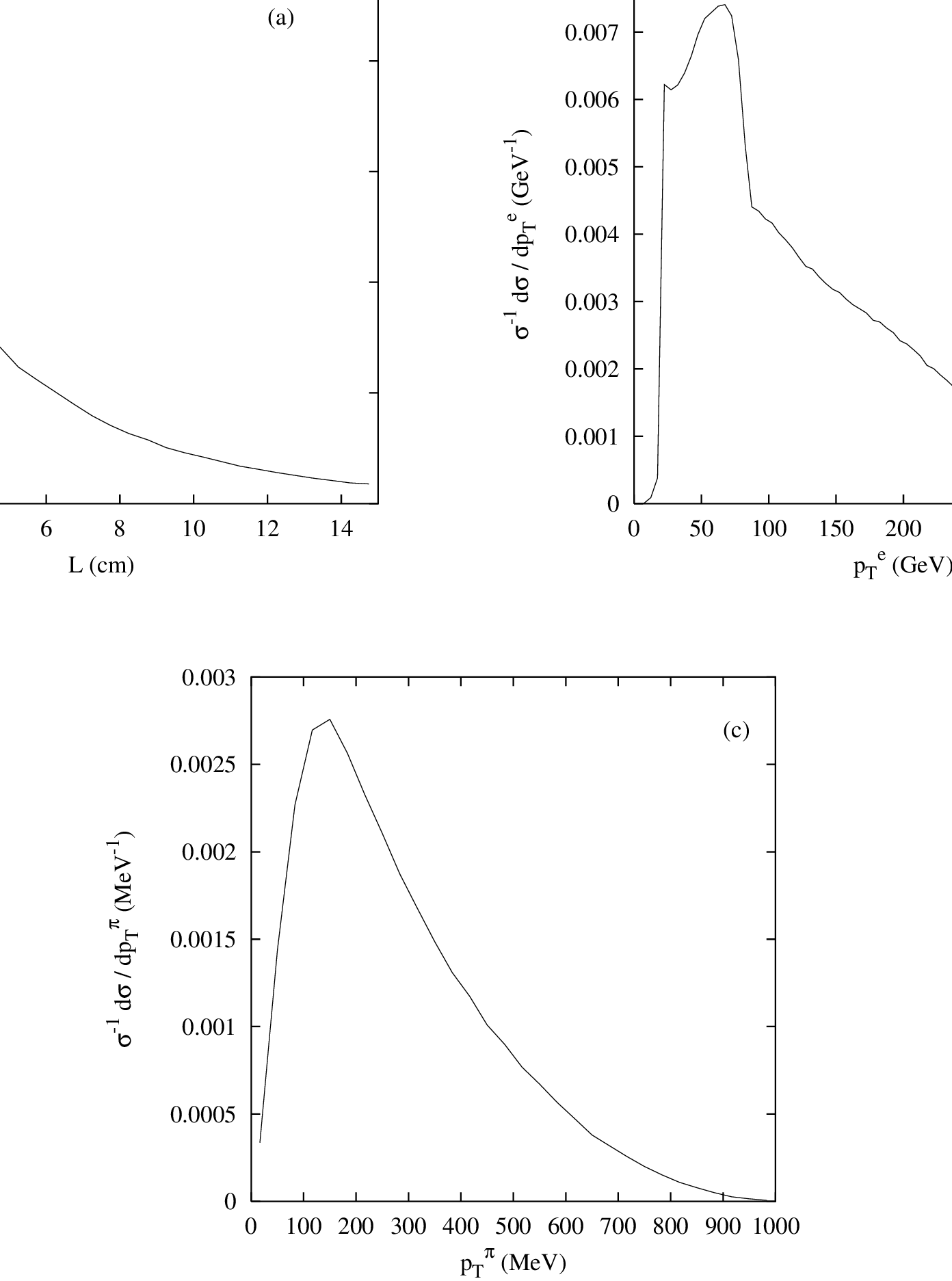,width=2.0in}}
\vspace*{8pt}
\caption{Normalized kinematic distributions of decay products:
$(a)$ decay length of the lighter chargino,
$(b)$ $p_T$ of charged lepton,
and $(c)$ $p_T$ of the charged pion arising from
$e^\pm + \pi^\mp + \mET$ signal for spectrum A.
The AMSB input parameters are $m_{3/2}=44$~TeV, $\tan\beta = 30$,
$\mu > 0$ and $m_0 = 410$ GeV.}
\end{figure}

The kinematic distributions of the final state particles for the
$e^\pm + \pi^\mp + \mET$ signal have been studied for the following
sample point in the AMSB parameter space corresponding to Spectrum A:
$m_{3/2}=44$~TeV, $\tan\beta=30$, $\mu > 0$ and $m_0 = 410$~GeV. These
are shown in Fig.1. The detailed discussions regarding these
distributions can be found in Ref. \cite{ghosh}. For these values of 
AMSB input parameters, $\Delta M = 198$ MeV. 

Before going to the next section let us also mention that the
possibility of observing same sign dilepton signals induced by the 
$\sneu\sneu^*$ mixing in the AMSB model (where $\tilde\tau$ is the LSP) 
has been discussed in Ref.\cite{choi} and AMSB models with bilinear 
R-parity violation have been shown to have testable signatures at a 
high energy $e^+e^-$ linear collider from the associated production of 
charged higgs and sleptons.\cite{diaz-rp}  

\section{Signals of AMSB in $e^- \gamma$ and $\gamma\gamma$ Collider} 

In Ref. \cite{e-gamma}, the process $e^- \gamma \r \sneu \Cm1$ was 
considered to look for signals of AMSB. 
Once produced, the sneutrino may decay into either an ($e^- \tilde
\chi_1^+$) pair or a ($\nu \tilde \chi_1^0$) pair. Concentrating upon 
the former, we are left with a fast $e^-$ (which serves as the trigger),
two heavily ionizing charged tracks coming from the long-lived 
$\C1pm$ and/or two visible soft pions with opposite 
charges and a large missing 
transverse momentum ($\mpT$). This is a very unique and distinct
signature of anomaly mediated supersymmetry breaking and does not
readily arise in either of mSUGRA or GMSB scenarios.

It turns out that, for most of the mAMSB parameter space,
the two charged pions are well separated from each
other. The neutralinos, on the other hand, escape detection,
thereby giving rise to an imbalance in momentum.

The signal, then is,
\be
e^-\gamma \ra \tilde\nu_e \tilde\chi^-_1 \ra e^- +\pi^+\pi^- + \mpT \ ,
\ee
with the energetic electron serving as the trigger for the event.
The relatively small decay width of the charginos is manifested
in heavily ionizing charged tracks (one for each chargino)
terminating inside the detector after traversing a macroscopic distance
and ending in a soft pion (with $p_T > 200$~MeV) in the Silicon Vertex
Detector (SVD) located very close to the beam pipe.
The details of the SM backgrounds, selection cuts and the kinematic
distributions can be found in Ref. \cite{e-gamma}. 

\subsection{Signal strength and the parameter space}
Let us now look into the total signal strength as a function of the
parameters involved. We shall restrict ourselves to two discrete values
of $\tan \beta$ while allowing $m_0$ and $m_{3/2}$ to vary freely
modulo the experimental constraints. As for the beam polarization, we
make a particular choice, namely $P_L = +1, P_{b} = P_{e^-} = -0.8$. 

In Fig.2, the results have been displayed for a machine
operating at $\sqrt{s}_{ee} = 500 \gev$, in the form of scatter
plots for the cross section, after imposing the cuts,
in the plane spanned by $m_0$ and $m_{3/2}$.
In each of the individual graphs, the region marked
by $X$ corresponds to a chargino mass of less than 86 GeV, thereby
falling foul of the ALEPH bound.\cite{aleph} The region $Y$, on the
other hand, would correspond to the $\tilde \tau_1$ being the LSP, a
possibility not allowed phenomenologically if $R$-parity is to 
be conserved.

\begin{figure}[th]
\vspace*{18pt}
\centerline{\psfig{file=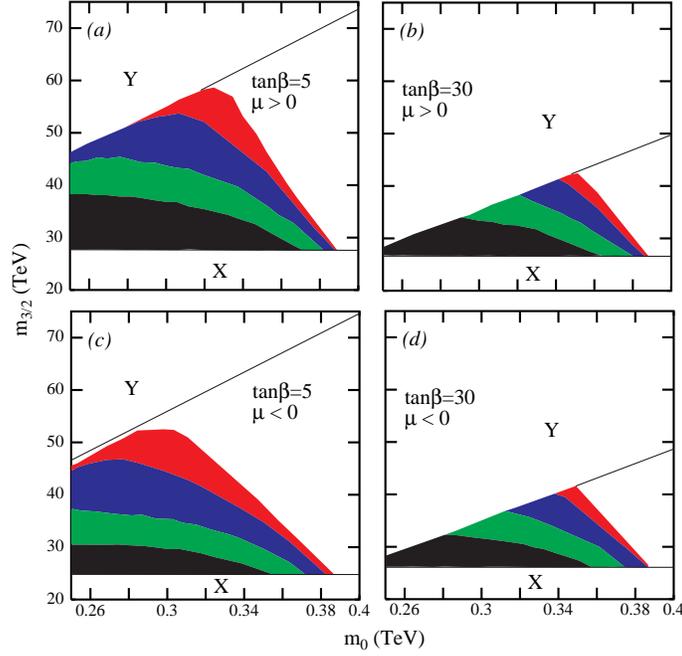,width=2.0in}}
\vspace*{8pt}
\caption{Scatter plots for the signal cross-section (fb) in the
         $m_0-m_{3/2}$ plane for a machine operating
         at $\sqrt{s}_{ee} = 500\gev$ and $P_L = +1, P_{b} = P_{e^-} 
         = -0.8$. The values of $tan\beta$ and $ sgn(\mu)$
         are as indicated. The regions marked by
        $X$ are ruled out by the experimental lower limit on the
        chargino mass, while those marked by $Y$ are ruled out by the 
         requirement of $\N10$ being the LSP.
         In each panel, the top three shaded
         regions correspond to cross section ranges of
         $[(0.1-5),~(5-50),~(50-150)]$. The lowermost region
         corresponds to $(150-470)$ in ${(a)}$,
         $(150-390)$ in ${(b)}$, $(150-335)$ in ${(c)}$
         and $(150-350)$ in ${(d)}$ respectively. This figure is taken
from \protect\cite{e-gamma}.}
\end{figure}

It is clear from Fig.2 that, for the low $tan\beta $
case, an experiment such as this can easily explore
$m_{3/2}$ as high as $60 (50)$~TeV
for negative (positive) $\mu$.
For $tan\beta = 30$, on the other hand, the maximal reach in $m_{3/2}$
is approximately 40 TeV irrespective of $sgn(\mu)$.
Similarly, the reach in $m_0$ has little dependence on either of
$tan\beta$ and $sgn(\mu)$.
Finally, it must be borne in mind that Fig.2
has been drawn keeping in mind a moderate luminosity ($\lsim 100 \fb^{-1}$).
A significantly larger integrated  luminosity would, of course, allow
one to explore beyond the topmost shaded band. A similar plot for a 
machine opearting at $\sqrt{s}_{ee} = 1 \tev$ instead can be seen in
Ref. \cite{e-gamma}. The features are very similar, with the obvious 
enhancement in the reach.

\subsubsection{Parameter determination}

The possibility of determining the masses of the chargino and the
sneutrino has also been investigated at an $e \gamma$ collider.
Since the particles produced were the sneutrino and the chargino
and the sneutrino subsequently decayed into a similar chargino and
an electron, it is easy to see that the energy of the decay electron is
strictly confined\cite{e-gamma} within the interval
\begin{equation}
{m_{\tilde \nu}^2-m_{\tilde\chi_1^+}^2
\over
2\left(
E_{\tilde \nu}^{\rm max} + k_{\tilde \nu}^{\rm max}
\right)}
\leq E^e \leq
{m_{\tilde \nu}^2-m_{\tilde\chi_1^+}^2
\over
2\left(
E_{\tilde \nu}^{\rm max} - k_{\tilde \nu}^{\rm max}
\right)}
\ ,
\label{e_elec}
\end{equation}
where $E_{\tilde \nu}^{\rm max}$ is the maximum possible energy that
the intermediate sneutrino may have carried, viz.
\begin{eqnarray}
E_{\tilde \nu}^{\rm max}
=
{1 \over 4y_{\rm max}\sqrt{s}}
&\Biggl[&
(1+y_{\rm max}) \left(y_{\rm max}s+m_{\tilde
\nu}^2-m_{\tilde\chi_1^+}^2\right)
\nonumber\\
&+&
(1-y_{\rm max})
        \sqrt{ \left(y_{\rm max}s+m_{\tilde
\nu}^2-m_{\tilde\chi_1^+}^2\right)^2                - 4y_{\rm
max}sm_{\tilde \nu}^2 }
\quad\Biggr] \ ,
\label{e_sneu}
\end{eqnarray}
and $k_{\tilde \nu}^{\rm max}$ is the corresponding momentum.
$y_{\rm max}$ is the maximum value of the fraction of the $e^\pm$ energy
that the reflected photon beam carries off.\cite{e-gamma}

\begin{figure}[th]
\vspace*{-1in}
\centerline{\psfig{file=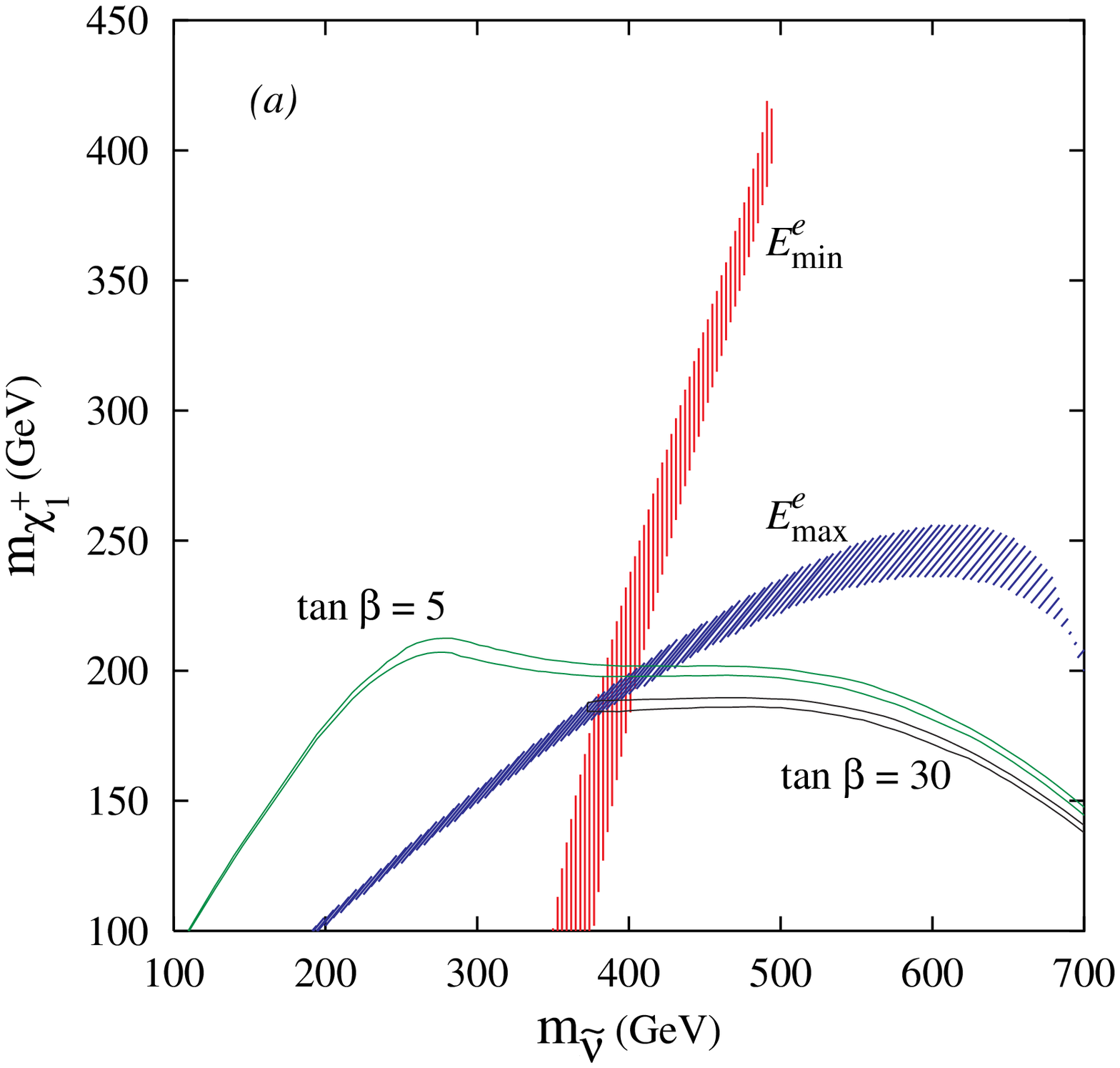,width=2.0in}
\psfig{file=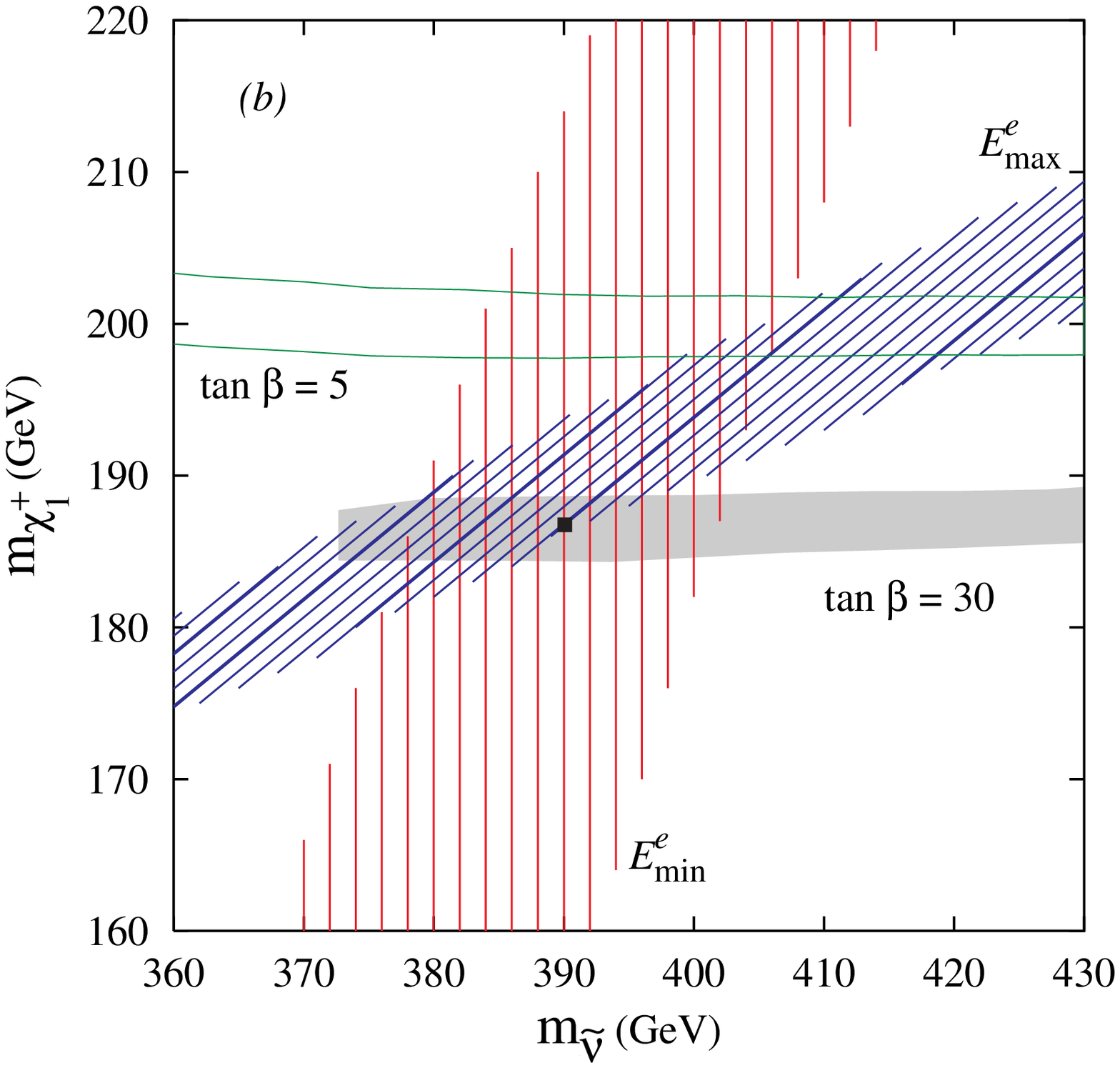,width=2.0in}}
\vspace*{8pt}
\caption[]{{\em (a)} The determination of the sneutrino and chargino
       masses from a measurement of the endpoints of the electron energy
       spectrum (eqn.\protect\ref{e_elec}). The width of the bands
correspond
        to the error bars in measuring the 
endpoints.\protect\cite{e-gamma} For points within
        the two horizontal bands, the resultant cross section would
agree
        with the measured one to within $1 \sigma$.
        {\em (b)} The overlap region has been shown on an expanded
scale.
        The dark square denotes the reference point in the parameter
        space (pt. {\bf B} in the text). This figure is taken from 
\protect\cite{e-gamma}.
        }
\end{figure}

As $\sqrt{s}$ and $y_{\rm max}$ are both known, it follows that an
accurate
measurement of $E^e_{\rm max}$ and  $E^e_{\rm min}$ would allow us to
determine both the masses. A very precise measurement of the endpoints
is unlikely, though. The measurement errors have been discussed in Ref.
\cite{e-gamma}. In Fig. 3{\em a}, we display the two bands obtained from
the measurement of the two electron energy endpoints corresponding to
a particular point in the parameter space ($m_0 = 450\gev, m_{3/2} =
55\tev, \tan\beta = 30$ and
$\mu > 0$\\
leading to $(m_{\Cp1}, m_{\N10}, m_{\tilde \nu})
                = (186.42, 186.23, 390.07) \gev$) which is denoted as
point {\bf B} in Ref. \cite{e-gamma}.  
The error on the sneutrino mass, thus determined, is roughly 12 GeV, while 
that on the chargino mass is roughly 6 GeV. Moreover, the errors are quite
correlated (see Fig.~3{\em b}, which displays the
region of interest on an expanded scale). 

Having determined the masses, it is now of interest to measure the
remaining parameter, namely $\tan \beta$. Clearly, kinematical
distributions
are essentially independent of this quantity and one should rather
consider
cross section measurement. 
The corresponding error in the cross-section
measurement (after imposing the cuts, naturally) is easily determined
on application of Poisson (or Gaussian) statistics.
Armed with this, and for a given value of
$\tan \beta$, one could easily determine the part of the
$m_{\tilde \nu}$--$m_{\tilde \chi_1^+}$ space that would be consistent
with the measured cross-section. In Figs.~3, we display
these constraints for two particular values of the ratio $\tan \beta$.
Note that, while some
resolution is possible, such experiments are not overly sensitive to
this
parameter. It is possible that significant improvement would occur once
other
production processes are considered. 

\subsection{AMSB signals in $\gamma \gamma$ collider}

Another possibility of observing AMSB signal is to look for chargino
pair production in association with a photon in $\gamma\gamma$
colisions. The production process is solely goverened by electromagnetic
interactions. The details of the cross sections, selection criteria and
background elimination can be found in Ref. \cite{gamma_gamma}. Assuming
an integrated luminosity of 100 $\fb^{-1}$, a chargino mass (considering
that the macroscopic track length cannot been seen and the soft charged
pions are not observed) up to about 165-170 $\gev$ can be detected at a
linear collider operating at $\sqrt s$ = 500 $\gev$ and the limit can go
up to about 370 $\gev$ for a 1 TeV collider. Another important feature
of this signal is that just by event counting itself one can have a
fairly good determination of $M_{\C1pm}$. 

Before we conclude, it may be relevant to discuss the ways of looking at
AMSB signals in the context of the Tevatron or the LHC. It has been
shown in Ref.\cite{feng} that at the Tevatron one could possibly detect
these charginos traveling macroscopic distances and the events can be
triggered on high $p_T$ monojets. Production and subsequent decays of
other sparticles have also been suggested for triggering on the 
events.\cite{wells} At the LHC, one could look at leptons + jets + charged
track + missing energy final states from the sparticle 
cascades\cite{paige,tata,barr} or forward jets + missing energy or one 
lepton + forward jets + missing energy from gauge boson 
fusion.\cite{konar,datta-huitu} As we have emphasized in this review 
the soft charged pions might play an important role in looking for 
signals of AMSB in linear colliders. In a hadronic environment it is 
almost impossible to look for these soft charged pions with 
characteristic momentum distribution. Thus, in some sense the searches
carried out in linear colliders along the direction discussed in this
review can be complementary to the searches in the hadron colliders.
Another important aspect is the measurement of the chargino and slepton
masses from the kinematics as well as from the cross-section
measurements which can possibly identify the underlying model. These
measurements may not be very reliable at a hadron collider. 

\section{Concluding remarks}

Anomaly mediated supersymmetry breaking is an interesting way of
generating slepton and gaugino masses. At low energies it is beset with
tachyonic sleptons. This problem is solved in various ways; here we  
discuss the minimal model where a constant term $m^2_0$ is added to all 
the squared masses of the scalars. Many interesting signatures can be 
seen in future linear colliders. One interesting feature of AMSB models 
is that the lighter chargino is winolike (nearly mass-degenerate with 
the winolike lightest neutralino) and its dominant decay mode is 
$\C1pm \rightarrow \N10$ + soft $\pi^\pm$. Fast $e^-$ or ISR photon 
trigger, heavily ionizing charged track from the slow decay of the 
lighter chargino and/or soft charged pion with impact parameter 
resolved can be a distinct possibility in an $e^+e^-$ linear collider. 
In an $e^-\gamma$ collider one could observe two soft charged pions and 
two heavily ionizing charged tracks with an $e^-$ trigger whereas in 
a $\gamma\gamma$ collider one can observe a single photon plus 
missing energy signal as a test of AMSB. A measurement of the 
fundamental supersymmetry breaking parameters could also be possible. 

\section*{Acknowledgments}

I am grateful to U. Chattopadhyay, D. Choudhury, E. Gabrielli, 
D.K. Ghosh, K. Huitu, A. Kundu and P. Roy for collaboration in various
works summarised in this review. I would like to thank the Lady Davis
Fellowship Trust in Technion for financial support while this work was
in progress. Thanks are due to A. Datta and B. Mukhopadhyaya for very
useful discussions. This work was supported by the Academy of Finland
(project number 48787).  

\section*{References}

\vspace*{6pt}

\def\pr#1,#2 #3 { {Phys.~Rev.}        ~{\bf #1},  #2 (19#3) }
\def\prd#1,#2 #3{ { Phys.~Rev.}       ~{D \bf #1}, #2 (19#3) }
\def\pprd#1,#2 #3{ { Phys.~Rev.}      ~{D \bf #1}, #2 (20#3) }
\def\prl#1,#2 #3{ { Phys.~Rev.~Lett.}  ~{\bf #1},  #2 (19#3) }
\def\pprl#1,#2 #3{ {Phys. Rev. Lett.}   {\bf #1},  #2 (20#3)}
\def\plb#1,#2 #3{ { Phys.~Lett.}       ~{\bf B#1}, #2 (19#3) }
\def\pplb#1,#2 #3{ {Phys. Lett.}        {\bf B#1}, #2 (20#3)}
\def\npb#1,#2 #3{ { Nucl.~Phys.}       ~{\bf B#1}, #2 (19#3) }
\def\pnpb#1,#2 #3{ {Nucl. Phys.}        {\bf B#1}, #2 (20#3)}
\def\prp#1,#2 #3{ { Phys.~Rep.}       ~{\bf #1},  #2 (19#3) }
\def\zpc#1,#2 #3{ { Z.~Phys.}          ~{\bf C#1}, #2 (19#3) }
\def\epj#1,#2 #3{ { Eur.~Phys.~J.}     ~{\bf C#1}, #2 (19#3) }
\def\mpl#1,#2 #3{ { Mod.~Phys.~Lett.}  ~{\bf A#1}, #2 (19#3) }
\def\ijmp#1,#2 #3{{ Int.~J.~Mod.~Phys.}~{\bf A#1}, #2 (19#3) }
\def\ptp#1,#2 #3{ { Prog.~Theor.~Phys.}~{\bf #1},  #2 (19#3) }
\def\jhep#1, #2 #3{ {J. High Energy Phys.} {\bf #1}, #2 (19#3)}
\def\pjhep#1, #2 #3{ {J. High Energy Phys.} {\bf #1}, #2 (20#3)}

\end{document}